\documentclass[10pt,journal,final,twocolumn]{IEEEtran}

\usepackage{graphicx}
\usepackage{subfigure}
\usepackage{epstopdf}
\usepackage{epsfig}
\usepackage[cmex10]{amsmath}
\usepackage{amssymb}
\usepackage{times}
\usepackage{ntheorem}
\usepackage{pifont}
\usepackage{stfloats}
\usepackage{bm}
\usepackage{color}
\usepackage{makecell}
\usepackage{array}
\usepackage{algorithm}
\usepackage{algorithmic}
\usepackage{multirow}

\begin{document}
%
\title{Deep CNN-Based Channel Estimation for mmWave Massive MIMO Systems}

\author{\IEEEauthorblockN{Peihao~Dong,~\IEEEmembership{Student Member,~IEEE},
Hua Zhang,~\IEEEmembership{Member,~IEEE}, Geoffrey Ye Li,~\IEEEmembership{Fellow,~IEEE},\\ Ivan Sim\~{o}es Gaspar, and Navid NaderiAlizadeh
}

\vspace{-0.3cm}
\thanks{

P. Dong and H. Zhang are with the National Mobile Communications Research Laboratory, Southeast University, Nanjing 210096, China (e-mail: phdong@seu.edu.cn; huazhang@seu.edu.cn).

G. Y. Li is with the School of Electrical and Computer Engineering, Georgia Institute of Technology, Atlanta, GA 30332 USA (e-mail: liye@ece.gatech.edu).

I. Gaspar and N. NaderiAlizadeh are with Intel Corporation, Santa Clara, CA 95054 USA (e-mail: ivan.simoes.gaspar@intel.com; navid.naderializadeh@intel.com)
}
\vspace{-0.2cm}}

\IEEEtitleabstractindextext{%
\begin{abstract}
For millimeter wave (mmWave) massive multiple-input multiple-output (MIMO) systems, hybrid processing architecture is usually used to reduce the complexity and cost, which poses a very challenging issue in channel estimation. In this paper, deep convolutional neural network (CNN) is employed to address this problem. We first propose a spatial-frequency CNN (SF-CNN) based channel estimation exploiting both the spatial and frequency correlation, where the corrupted channel matrices at adjacent subcarriers are input into the CNN simultaneously. Then, exploiting the temporal correlation in time-varying channels, a spatial-frequency-temporal CNN (SFT-CNN) based approach is developed to further improve the accuracy. Moreover, we design a spatial pilot-reduced CNN (SPR-CNN) to save spatial pilot overhead for channel estimation, where channels in several successive coherence intervals are grouped and estimated by a channel estimation unit with memory. Numerical results show that the proposed SF-CNN and SFT-CNN based approaches outperform the non-ideal minimum mean-squared error (MMSE) estimator but with reduced complexity, and achieve the performance close to the ideal MMSE estimator that is very difficult to be implemented in practical situations. They are also robust to different propagation scenarios. The SPR-CNN based approach achieves comparable performance to SF-CNN and SFT-CNN based approaches while only requires about one third of spatial pilot overhead at the cost of complexity. Our work clearly shows that deep CNN can efficiently exploit channel correlation to improve the estimation performance for mmWave massive MIMO systems.
\end{abstract}

}

\maketitle

\IEEEdisplaynontitleabstractindextext

%
\IEEEpeerreviewmaketitle

\vspace{-0.2cm}
\section{Introduction}

Millimeter wave (mmWave) communications can meet the high data rate demand due to its large bandwidth. Its high propagation loss can be well compensated by using massive multiple-input multiple-output (MIMO) technique \cite{A. L. Swindlehurst}$-$\cite{M. Jian}. However, Due to the limited physical space with closely placed antennas and prohibitive power consumption in mmWave massive MIMO systems, it is difficult to equip a dedicated radio frequency (RF) chain for each antenna. To reduce complexity and cost, phase shifter based two-stage architecture, usually called hybrid architecture, is widely used at both the transmitter and the receiver to connect a large number of antennas with much fewer RF chains \cite{R. W. Heath Jr.}, \cite{L. Liang}.

For mmWave massive MIMO systems with the hybrid architecture, channel estimation is a challenging problem. In \cite{A. Alkhateeb}, a hierarchical multi-resolution codebook has been designed, based on which an adaptive channel estimation algorithm has been developed by exploiting the channel sparsity. In \cite{Z. Gao}, the structured sparsity in angle domain has been utilized to estimate the wideband channel for multi-user mmWave massive MIMO uplink. In \cite{Y. Wang}, a channel estimation approach has been developed for mmWave massive MIMO orthogonal frequency division multiplexing (OFDM) systems with low-precision analog-to-digital converters (ADCs). For the mmWave MIMO-OFDM downlink, a channel parameter estimation for the angles, time delays, and fading coefficients has been proposed in \cite{Z. Zhou} resorting to the low-rank tensor decomposition. Instead of estimating the mmWave MIMO channel directly, the method for singular subspace estimation has been proposed in \cite{H. Ghauch}, based on which a subspace decomposition algorithm has been further developed to design the hybrid analog-digital architecture.

Compared to the conventional methods, machine learning (ML) is more powerful to uncover the inherent characteristics inside data/signals collected in an end-to-end manner and thus can achieve better performance when addressing various problems in wireless communications \cite{Z. Qin}. In \cite{H. Ye}, deep learning (DL) has been successfully used in joint channel estimation and signal detection of OFDM systems with interference and non-linear distortions. In \cite{H. He}, iterative channel estimation has been proposed for the lens mmWave massive MIMO systems, where denoising neural network (NN) is used in each iteration to update the estimated channel. To reduce the CSI feedback overhead of the frequency duplex division (FDD) massive MIMO system, DL has been employed in \cite{C.-K. Wen} to compress the channel into a low dimensional codeword and then to perform recovery with high accuracy. Exploiting temporal correlation of the channel, long short-term memory (LSTM) based deep NN (DNN) has been introduced in \cite{T. Wang} to develop a more efficient channel compression and recovery method for the CSI feedback. More related research can be also found in \cite{H. He_b} and the references therein. In \cite{J.-M. Kang}, DL has been applied to estimate channels in wireless power transfer systems, which outperforms the conventional scheme in terms of both estimation accuracy and harvested energy. In \cite{P. Dong}, supervised learning algorithms have been used to acquire the downlink CSI for FDD massive MIMO systems with reduced overheads for pilot and CSI feedback. In \cite{Y.-S. Jeon}, the supervised learning has been exploited to perform blind detection for massive MIMO systems with low-precision ADCs.

The conventional channel estimation methods usually perform unsatisfactorily in the more practical complicated channel model and also suffer from high complexity. In
contrast, the deep convolutional NN (CNN) is more capable to extract the inherent characteristics underlying the channel matrix from the large amount of data and provides the potential to estimate the channel more accurately with lower complexity by using the efficient parallel computing methods. In this paper, we use the deep CNN to address channel estimation for mmWave massive MIMO-OFDM systems. To exploit the correlation among channels at adjacent subcarriers in OFDM, we first propose a spatial-frequency CNN (SF-CNN) based channel estimation, where the tentatively estimated channel matrices at adjacent subcarriers are input into the CNN simultaneously \cite{P. Dong_a}. To further exploit the temporal correlation, a spatial-frequency-temporal CNN (SFT-CNN) based channel estimation is developed, where the channel information of the previous coherence interval is utilized when estimating the channel matrices of the current coherence interval. The SFT-CNN based approach incorporates all types of channel correlation in a simple way and yields remarkable performance gains that can be used to significantly save the spatial pilot overhead due to large-scale arrays. Therefore, we propose a spatial pilot-reduced CNN (SPR-CNN) based channel estimation, where channels in several successive coherence intervals are grouped and estimated by a channel estimation unit (CEU) with memory. From the numerical results, the proposed SF-CNN and SFT-CNN based approaches outperform the non-ideal minimum mean-squared error (MMSE) estimator and achieve the performance very close to the ideal MMSE estimator that is very difficult to be implemented in practical systems. They are also with lower complexity than the MMSE estimator and exhibit the unique robustness to maintain the fairly good performance when facing different channel statistics. The SPR-CNN based approach achieves comparable performance to SF-CNN and SFT-CNN based approaches by using only about one third of spatial pilot overhead and moderately increased complexity.

The rest of the paper is organized as follows. Section II describes the considered mmWave massive MIMO system, followed by the proposed SF-CNN based channel estimation in Section III. Section IV further develops the SFT-CNN and SPR-CNN based channel estimation. Numerical results are provided in Section V and finally Section VI gives concluding remarks.

\emph{Notations}: In this paper, we use upper and lower case boldface letters to denote matrices and vectors, respectively. $\lVert\cdot\rVert_{F}$, $(\cdot)^T$, $(\cdot)^H$, $(\cdot)^{-1}$, and $\mathbb{E}\{\cdot\}$ represent the Frobenius norm, transpose, conjugate transpose, inverse, and expectation, respectively. $\mathcal{CN}(\mu,\sigma^2)$ represents circular symmetric complex Gaussian distribution with mean $\mu$ and variance $\sigma^2$. $\delta(\cdot)$ and $\lceil\cdot\rceil$ denote the delta function and ceiling function. $\mathbb{N}_{+}$ denotes the set of all positive integers.

\vspace{-0.2cm}
\section{System Model}

\begin{figure*}[t]
\centering
\includegraphics[width=6.8in]{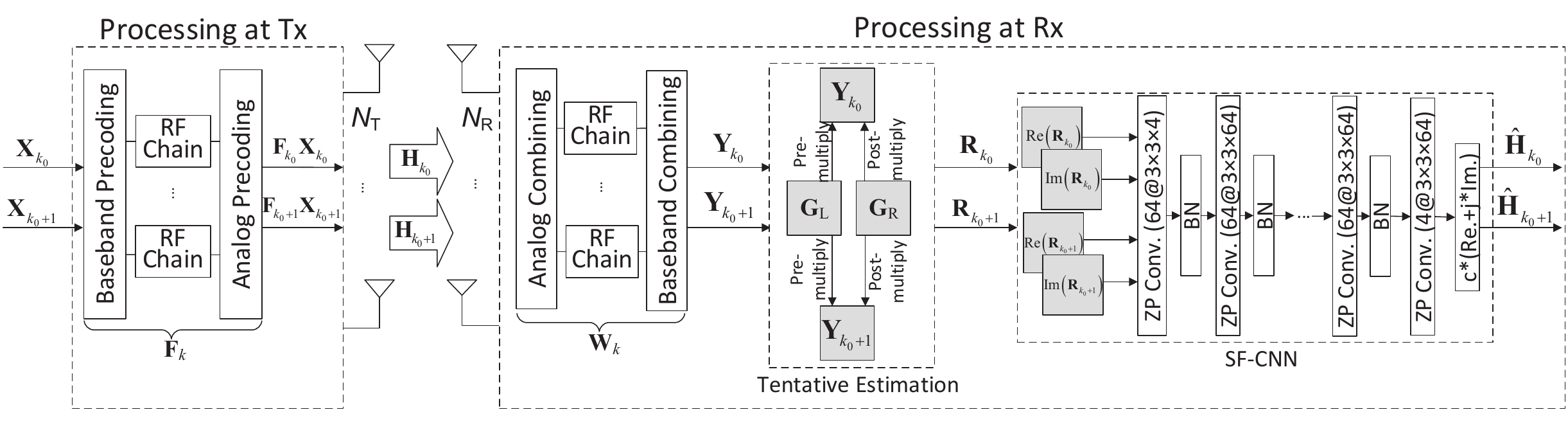}
\caption{SF-CNN based channel estimation.}\label{SF-CNN}
\end{figure*}

We consider a mmWave massive MIMO-OFDM system as in Fig.~\ref{SF-CNN}, where the transmitter is with $N_{\textrm{T}}$ antennas and $N_{\textrm{T}}^{\textrm{RF}}$ RF chains and the receiver is with $N_{\textrm{R}}$ antennas and $N_{\textrm{R}}^{\textrm{RF}}$ RF chains. Phase shifters are employed to connect a large number of antennas with a much fewer number of RF chains at both the transmitter and the receiver sides. We therefore assume $N_{\textrm{T}}\gg N_{\textrm{T}}^{\textrm{RF}}$ and $N_{\textrm{R}}\gg N_{\textrm{R}}^{\textrm{RF}}$.

According to \cite{Z. Gao}, the $N_{\textrm{R}}\times N_{\textrm{T}}$ channel matrix between the receiver and the transmitter in the delay domain is given by
\begin{equation}
\label{eqn_H_tau}
\mathbf{H}(\tau)=\sqrt{\frac{N_{\textrm{T}}N_{\textrm{R}}}{L}}\sum_{l=1}^L \alpha_{l}\delta(\tau-\tau_l)\mathbf{a}_{\textrm{R}}(\varphi_l)\mathbf{a}_{\textrm{T}}^{H}(\phi_l),
\end{equation}
where $L$ is the number of main paths, $\alpha_{l}\sim \mathcal{CN}(0, \sigma_{\alpha}^2)$ is the propagation gain of the $l$th path with $\sigma_{\alpha}^2$ being the average power gain, $\tau_l$ is the delay of the $l$th path, $\varphi_l$ and $\phi_l\in[0,2\pi]$ are the azimuth angles of arrival and departure (AoA/AoD) at the receiver and the transmitter, respectively. For uniform linear array (ULA), the corresponding response vectors can be expressed as
\begin{equation}
\label{eqn_au}
\mathbf{a}_{\textrm{R}}(\varphi_l)=\frac{1}{\sqrt{N_{\textrm{R}}}}\bigl[1,e^{-j2\pi\frac{d}{\lambda}\sin(\varphi_l)},\ldots,e^{-j2\pi\frac{d}{\lambda}(N_{\textrm{R}}-1)\sin(\varphi_l)}\bigr]^{T},
\end{equation}
\begin{equation}
\label{eqn_au}
\mathbf{a}_{\textrm{T}}(\phi_l)=\frac{1}{\sqrt{N_{\textrm{T}}}}\bigl[1,e^{-j2\pi\frac{d}{\lambda}\sin(\phi_l)},\ldots,e^{-j2\pi\frac{d}{\lambda}(N_{\textrm{T}}-1)\sin(\phi_l)}\bigr]^{T},
\end{equation}
where $d$ and $\lambda$ denote the distance between the adjacent antennas and carrier wavelength, respectively.

According to the channel model in (\ref{eqn_H_tau}), the frequency domain channel of the $k$th subcarrier in OFDM is given by
\begin{equation}
\label{eqn_H_k}
\mathbf{H}_k=\sqrt{\frac{N_{\textrm{T}}N_{\textrm{R}}}{L}}\sum_{l=1}^L \alpha_{l}e^{-j2\pi\tau_l f_s\frac{k}{K}}\mathbf{a}_{\textrm{R}}(\varphi_l)\mathbf{a}_{\textrm{T}}^{H}(\phi_l),
\end{equation}
where $f_s$ denotes the sampling rate and $K$ is the number of OFDM subcarriers.

To estimate $\mathbf{H}_k$, the transmitter activates only one RF chains to transmit the pilot on one beam in each channel use while the receiver combines the received pilot by using all RF chains associated with different beams. In more detail, the transmitter transmits pilots, $x_{k,u}$, using $M_{\textrm{T}}$ beamforming vectors, $\mathbf{f}_{k,u}\in\mathbb{C}^{N_{\textrm{T}}\times 1}$, $u=1,\ldots,M_{\textrm{T}}$. For the transmit pilot signal corresponding to each beamforming vector, $\mathbf{f}_{k,u}$, the receiver employs $M_{\textrm{R}}$ combining vectors, $\mathbf{w}_{k,v}\in\mathbb{C}^{N_{\textrm{R}}\times 1}$, $v=1,\ldots,M_{\textrm{R}}$, to respectively process it. Since the receiver is equipped with $N_{\textrm{R}}^{\textrm{RF}}(<M_{\textrm{R}})$ RF chains, it can only use $N_{\textrm{R}}^{\textrm{RF}}$ combining vectors in a channel use. Then, if the receiver uses all $M_{\textrm{R}}$ combining vectors to process a beamforming vector carrying pilot, the required channel uses will be $\left\lceil\frac{M_{\textrm{R}}}{N_{\textrm{R}}^{\textrm{RF}}}\right\rceil$. So the total channel uses for processing all beamforming vectors will be $M_{\textrm{T}}\left\lceil\frac{M_{\textrm{R}}}{N_{\textrm{R}}^{\textrm{RF}}}\right\rceil$.\footnote{This pilot transmission process can capture the main paths in mmWave channels. Although simultaneously activating multiple RF chains with different beams at the transmitter can accelerate the pilot transmission process, it fails to capture the main paths and leads to poor channel estimation performance. Thus in lots of related work, only one RF chain is activated to transmit the pilot on one beam at the transmitter during each channel use to guarantee the performance of channel estimation algorithms \cite{A. Alkhateeb}, \cite{Y. Wang}. Similarly, in this paper, the pilot transmission mode that we consider also facilitates the CNN based channel estimation to achieve very good accuracy.}
Then the pilot signal matrix associated with the $k$th subcarrier at the baseband of the receiver can be written as
\begin{equation}
\label{eqn_Y_k}
\mathbf{Y}_k=\mathbf{W}_k^H\mathbf{H}_k\mathbf{F}_k\mathbf{X}_k+\tilde{\mathbf{N}}_k,
\end{equation}
where $\mathbf{W}_k=[\mathbf{w}_{k,1},\ldots,\mathbf{w}_{k,M_{\textrm{R}}}]$ and $\mathbf{F}_k=[\mathbf{f}_{k,1},\ldots,\mathbf{f}_{k,M_{\textrm{T}}}]$ are combining matrix and beamforming matrix, respectively, $\mathbf{X}_k$ is an $M_{\textrm{T}}\times M_{\textrm{T}}$ diagonal matrix with its $u$th diagonal element being $x_{k,u}$. $\tilde{\mathbf{N}}_k=\mathbf{W}_k^H\mathbf{N}_k$ denotes the effective noise after combining and $\mathbf{N}_k$ is additive white Gaussian noise (AWGN) with $\mathcal{CN}(0,1)$ elements before combining.

We consider the pilot insertion in both frequency and time domain. Specifically, adjacent $Q$ ($Q\geq 2$) subcarriers respectively place pilots with the same length at the beginning of a coherence interval to form a pilot subcarrier block and the rest of time slots in each coherence interval are used for data transmission. Two pilot subcarrier blocks are separated by $Q_{\textrm{d}}$ ($Q_{\textrm{d}}\geq 0$) subcarriers dedicated to data transmission. Pilots are utilized to estimate the channels of corresponding time-frequency positions. Based on the estimated channels, interpolation can be applied to get the channels at the positions without pilots. It is clear that the accuracy of interpolation is dependent on the accuracy of the estimated channels and the variation of channels. Therefore, we will focus on improving the accuracy of the pilot based channel estimation in this paper so that more reliable reference values can be provided for interpolation.

\section{SF-CNN based Channel Estimation}

In this section, we first elaborate the SF-CNN based channel estimation, including an overview of the proposed approach, the offline training of SF-CNN, and the online deployment. Then the computational complexity for the online estimation is analyzed.

\vspace{-0.2cm}
\subsection{Algorithm Description}

\emph{1) Channel Estimation Procedure:} Fig.~\ref{SF-CNN} illustrates the channel estimation procedure for adjacent $Q$ ($=2$) subcarriers, $k_{0}$ and $k_{0}+1$, to simplify our discussion even if it is trivial to extend to the case with $Q>2$. Without loss of generality, we assume the worst case that $\mathbf{W}_k=\mathbf{W}$, $\mathbf{F}_k=\mathbf{F}$, and $\mathbf{X}_k=\sqrt{P}\mathbf{I}$ for $k\in\left\{k_{0},k_{0}+1\right\}$, where $P$ denotes the transmit power. The pilot signal matrix, $\mathbf{Y}_k$, becomes
\setlength{\arraycolsep}{0.0em}
\begin{eqnarray}
\label{eqn_simple_Yk}
\mathbf{Y}_k=\sqrt{P}\mathbf{W}^H\mathbf{H}_k\mathbf{F}+\tilde{\mathbf{N}}_k.
\end{eqnarray}

Then $\mathbf{Y}_k$ goes through the tentative estimation (TE) module, which uses two matrices, $\mathbf{G}_{\textrm{L}}$ and $\mathbf{G}_{\textrm{R}}$, to process $\mathbf{Y}_k$ and outputs a coarse estimation of $\mathbf{H}_k$, that is,
\setlength{\arraycolsep}{0.0em}
\begin{eqnarray}
\label{eqn_Rk}
\mathbf{R}_k=\mathbf{G}_{\textrm{L}}\mathbf{Y}_k\mathbf{G}_{\textrm{R}}=\sqrt{P}\mathbf{G}_{\textrm{L}}\mathbf{W}^H\mathbf{H}_k\mathbf{F}\mathbf{G}_{\textrm{R}} +\mathbf{G}_{\textrm{L}}\tilde{\mathbf{N}}_k\mathbf{G}_{\textrm{R}},
\end{eqnarray}
where
\setlength{\arraycolsep}{0.0em}
\begin{eqnarray}
\label{eqn_GL}
\mathbf{G}_{\textrm{L}}=
\begin{cases}
\mathbf{W}, & M_{\textrm{R}}<N_{\textrm{R}},\\
(\mathbf{W}\mathbf{W}^H)^{-1}\mathbf{W}, & M_{\textrm{R}}\geq N_{\textrm{R}},
\end{cases}
\end{eqnarray}
and
\setlength{\arraycolsep}{0.0em}
\begin{eqnarray}
\label{eqn_GR}
\mathbf{G}_{\textrm{R}}=
\begin{cases}
\mathbf{F}^H, & M_{\textrm{T}}<N_{\textrm{T}},\\
\mathbf{F}^H(\mathbf{F}\mathbf{F}^H)^{-1}, & M_{\textrm{T}}\geq N_{\textrm{T}}.
\end{cases}
\end{eqnarray}

The tentatively estimated channel matrices $\mathbf{R}_{k_{0}}$ and $\mathbf{R}_{k_{0}+1}$ are then input into the SF-CNN simultaneously, which outputs the estimated channel matrices $\hat{\mathbf{H}}_{k_{0}}$ and $\hat{\mathbf{H}}_{k_{0}+1}$ through the mapping relationship
\setlength{\arraycolsep}{0.0em}
\begin{eqnarray}
\label{eqn_SFCNNmap}
\left\{\hat{\mathbf{H}}_{k_{0}}, \hat{\mathbf{H}}_{k_{0}+1}\right\}=f_{\Phi}\left(\mathbf{R}_{k_{0}},\mathbf{R}_{k_{0}+1};\Phi\right),
\end{eqnarray}
where $\Phi$ denotes the parameter set of the SF-CNN.

\emph{2) SF-CNN Offline Training:} For the proposed SF-CNN, the training set consisting of $N_{\textrm{tr}}$ samples is generated according to certain channel model in the simulation environment with $\left(\underline{\mathbf{R}}_i,\underline{\mathbf{H}}_i\right)$ denoting the $i$th sample, where $\underline{\mathbf{R}}_i$ is the input data and $\underline{\mathbf{H}}_i$ is the target data. $\underline{\mathbf{R}}_i\in\mathbb{C}^{N_{\textrm{R}}\times N_{\textrm{T}}\times 2}$ is a three-dimensional matrix composed of $\mathbf{R}_{k_{0}'}^{i},\mathbf{R}_{k_{0}'+1}^{i}\in\mathbb{C}^{N_{\textrm{R}}\times N_{\textrm{T}}}$, which are the tentatively estimated channel matrices at subcarrier $k_{0}'$ and $k_{0}'+1$ collected through (\ref{eqn_Rk}) with $k_{0}'\in\{1,2,\ldots,K-1\}$. $\underline{\mathbf{H}}_i\in\mathbb{C}^{N_{\textrm{R}}\times N_{\textrm{T}}\times 2}$ is also a three-dimensional matrix composed of $\frac{\mathbf{H}_{k_{0}'}^{i}}{c},\frac{\mathbf{H}_{k_{0}'+1}^{i}}{c}\in\mathbb{C}^{N_{\textrm{R}}\times N_{\textrm{T}}}$, where $\mathbf{H}_{k_{0}'}^{i}$ and $\mathbf{H}_{k_{0}'+1}^{i}$ are the corresponding true channel matrices. $c>0$ is a scaling constant to make the value range of the real and imaginary parts of all the target data, $\underline{\mathbf{H}}_i$, match the activation function applied in the output layer of the SF-CNN. Then $\underline{\mathbf{R}}_i$ is fed into the SF-CNN to approximate the corresponding scaled true channels $\underline{\mathbf{H}}_i$.

For the mmWave massive MIMO systems, we use $N_{\textrm{T}}=32$, $N_{\textrm{R}}=16$ as a typical example. As shown in Fig.~\ref{SF-CNN}, the SF-CNN receives the tentatively estimated complex channel matrices, $\mathbf{R}_{k_{0}'}^{i}\in\mathbb{C}^{16\times32}$ and $\mathbf{R}_{k_{0}'+1}^{i}\in\mathbb{C}^{16\times32}$, as the input and separates their real and imaginary parts so that four $16\times32$ real-valued matrices are obtained. In the subsequent convolutional layer, the four matrices are processed by $64$ $3\times3\times4$ convolutional filters with the rectified linear unit (ReLU) activation function to generate $64$ $16\times32$ real-valued matrices. Zero padding (ZP) is used when processing each feature matrix so that its dimension maintains unchanged after convolution. Then a batch normalization (BN) layer is added to avoid the gradient diffusion and overfitting. For the next eight convolutional layers, each uses $64$ $3\times3\times64$ filters to operate ZP convolution with the feature matrices passed by the previous layer and outputs $64$ $16\times32$ real-valued feature matrices. ReLU activation function is applied for these eight layers, each of which is followed by a BN layer. The output layer uses four $3\times3\times64$ convolutional filters to process the $64$ $16\times32$ real-valued feature matrices and obtains the estimated real and imaginary parts of the scaled channel matrices at the $k_{0}'$th and $(k_{0}'+1)$th subcarrier, i.e., $\frac{\textrm{Re}(\hat{\mathbf{H}}_{k_{0}'}^{i})}{c}$, $\frac{\textrm{Im}(\hat{\mathbf{H}}_{k_{0}'}^{i})}{c}$, $\frac{\textrm{Re}(\hat{\mathbf{H}}_{k_{0}'+1}^{i})}{c}$, and $\frac{\textrm{Im}(\hat{\mathbf{H}}_{k_{0}'+1}^{i})}{c}$. Hyperbolic tangent activation function is used in the output layer to map the output into interval $[-1,1]$. After scaling up and combining the corresponding real and imaginary parts, the $16\times32$ complex-valued estimated channel matrices, $\hat{\mathbf{H}}_{k_{0}'}^{i}$ and $\hat{\mathbf{H}}_{k_{0}'+1}^{i}$, are obtained. Table~\ref{table_SFCNN} lists the detailed architecture of the SF-CNN.

\begin{table}[!t]
\centering
\caption{Architecture of the SF-CNN}
\label{table_SFCNN}
\begin{tabular}{c|c|c|c|c}
\hline
 \makecell{Layer} & \makecell{Type} & \makecell{Number\\of filters} & \makecell{Size\\ of filters} & \makecell{Activation\\ function}\\
\hline
 1 & Input & - & - & - \\
\hline
 2$\sim$10 & Conv. & 64 & $3\times3$ & ReLU \\
\hline
 11 & Output & 2$Q$ & $3\times3$ & tanh \\
\hline
\end{tabular}
\end{table}

The objective of the offline training for the SF-CNN is to minimize the MSE loss function
\vspace{-0.1cm}
\begin{equation}
\begin{aligned}
\label{eqn_mse}
\textrm{MSE}_{\textrm{Loss}} =\frac{1}{N_{\textrm{tr}}c^2}\sum_{i=1}^{N_{\textrm{tr}}}\sum_{q=1}^{2}\left\|\mathbf{H}_{k_{0}'+q-1}^{i}-\hat{\mathbf{H}}_{k_{0}'+q-1}^{i}\right\|_{F}^2.
\end{aligned}
\end{equation}

The design of the SF-CNN architecture draws from CNN based image processing and considers the specific channel estimation task in this paper. The SF-CNN is used for channel denoising and thus we set the size of feature maps of each layer as $N_{\textrm{R}}\times N_{\textrm{T}}$. Nine convolutional hidden layers are used to fully uncover the inherent structure of the channel. According to \cite{K. Simonyan}, we adopt multiple convolutional filters with a very small size, i.e., $3\times 3$, to achieve good channel estimation performance with the low complexity. Based on our simulation trials, further increasing the number of convolutional layers or the number of convolutional filters does not bring major performance improvement but causes much higher complexity for CNN training and testing.

\emph{3) Online Deployment Issue:} After the offline training, the SF-CNN as well as the TE module will be deployed at the receiver to output the estimated channel matrices, $\hat{\mathbf{H}}_{k_{0}},\hat{\mathbf{H}}_{k_{0}+1},\ldots,\hat{\mathbf{H}}_{k_{0}+Q-1}$, by jointly processing the pilot matrices, $\mathbf{Y}_{k_{0}},\mathbf{Y}_{k_{0}+1},\ldots,\mathbf{Y}_{k_{0}+Q-1}$. If the actual channel model differs from what is used to generate the training set, a straightforward solution is fine-tuning. But it is difficult to collect the true channel and therefore the estimated channel is used instead. It is clear that using more power or longer pilot sequence makes the estimated channel closer to the true channel, which, however, increases the overhead for online fine-tuning. Fortunately, as shown by Fig.~\ref{SF_CNN_robustness} and Fig.~\ref{SF_CNN_robustness_diffpath} in Section V, the offline trained SF-CNN is quite robust to most of the new channel statistics that are not observed before. This implies that further online fine-tuning only provides marginal performance improvement and hence might not be necessary.

\subsection{Complexity Analysis}

In this subsection, we analyze the computational complexity of the proposed SF-CNN based channel estimation in testing stage and compare it with the non-ideal MMSE using estimated covariance matrix. The required number of floating point operations (FLOPs) is used as the metric.

For the proposed approach, the FLOPs come from the TE module processing in (\ref{eqn_Rk}) and SF-CNN. By assuming $M_{\textrm{T}}=N_{\textrm{T}}$ and $M_{\textrm{R}}=N_{\textrm{R}}$, the matrix product in (\ref{eqn_Rk}) requires FLOPs of $C_{\textrm{TE}}\sim\mathcal{O}(Q N_{\textrm{T}}N_{\textrm{R}}(N_{\textrm{T}}+N_{\textrm{R}}))$ \cite{R. Hunger}. According to \cite{K. He}, the required FLOPs of SF-CNN processing is $C_{\textrm{SF-CNN}}\sim\mathcal{O}\left(\sum_{l=1}^{L_{\textrm{c}}}M_{1,l}M_{2,l}F_{l}^2N_{l-1}N_{l}\right)$, where $L_{\textrm{c}}$ is number of convolutional layers, $M_{1,l}$ and $M_{2,l}$ denote the number of rows and columns of each feature map output by the $l$th layer, $F_{l}$ is the side length of the filters used by the $l$th layer, $N_{l-1}$ and $N_{l}$ denote the numbers of input and output feature maps of the $l$th layer. Specifically, these parameters are listed in Table~\ref{SFCNN_para} based on the SF-CNN offline training mentioned above. Then the computational complexity of the proposed SF-CNN based channel estimation is given by
\setlength{\arraycolsep}{0.0em}
\begin{eqnarray}
\label{eqn_complexity_CNN}
&&C_{\textrm{SF-CNN-CE}}\nonumber\\
&&\sim\mathcal{O}\left(Q N_{\textrm{T}}N_{\textrm{R}}(N_{\textrm{T}}+N_{\textrm{R}}) +N_{\textrm{T}}N_{\textrm{R}}\sum_{l=1}^{L_{\textrm{c}}}F_{l}^2N_{l-1}N_{l}\right)\!\!,\,\,\,
\end{eqnarray}

\begin{table}
  \centering
  \caption{SF-CNN Parameter Settings}
  \label{SFCNN_para}
  \begin{tabular}{c|c|c|c|c|c}
  \hline
  $l$ & $M_{1,l}$ & $M_{2,l}$ & $F_{l}$ & $N_{l-1}$ & $N_{l}$\\
  \hline
  $1$ & $16$ & $32$ & $3$ & $4$ & $64$\\
  \hline
  $2\sim9$ & $16$ & $32$ & $3$ & $64$ & $64$\\
  \hline
  $10$ & $16$ & $32$ & $3$ & $64$ & $4$\\
  \hline
  \end{tabular}
\end{table}

For the MMSE channel estimation, least-square (LS) channel estimation needs to be first performed causing FLOPs of $C_{\textrm{LS}}\sim\mathcal{O}(QN_{\textrm{T}}^2N_{\textrm{R}}^2)$. The channel covariance matrix is then calculated based on the LS channel estimation once per channel realization, which requires the computational complexity of $C_{\textrm{MMSE,}1}\sim\mathcal{O}(Q^2N_{\textrm{T}}^2N_{\textrm{R}}^2)$ if considering both spatial and frequency channel statistics. Finally, the LS channel estimation is refined by the covariance matrix and the corresponding FLOPs is $C_{\textrm{MMSE,}2}\sim\mathcal{O}(Q^3N_{\textrm{T}}^3N_{\textrm{R}}^3)$. Therefore, the overall computational complexity of MMSE is
\begin{equation}
\begin{aligned}
\label{eqn_complexity_MMSE}
C_{\textrm{MMSE}}\sim\mathcal{O}(Q^3N_{\textrm{T}}^3N_{\textrm{R}}^3).
\end{aligned}
\end{equation}

It is hard to compare $C_{\textrm{SF-CNN-CE}}$ with $C_{\textrm{MMSE}}$ straightforwardly in general since the former depends on $L_{\textrm{c}}$, $F_{l}$, $N_{l-1}$, and $N_{l}$ besides $Q$, $N_{\textrm{T}}$ and $N_{\textrm{R}}$. If $N_{\textrm{T}}=32$, $N_{\textrm{R}}=16$, $Q=2$ and other parameters for the SF-CNN are listed in Table~\ref{SFCNN_para}, the computational complexity of the proposed SF-CNN based approach is in the order of magnitude of $10^{8}$ while MMSE needs a higher complexity in the order of magnitude of $10^{9}$. In addition, the SF-CNN is able to run in a more efficient parallel manner and the runtime of a channel realization is only $1.47\times10^{-4}$ seconds by using NVIDIA GeForce GTX 1080 Ti GPU. By comparison, the MMSE consumes the time of about $6.14\times10^{-2}$ seconds per channel realization on the Intel(R) Core(TM) i7-3770 CPU.

\section{SFT-CNN and SPR-CNN based Channel Estimation}

\begin{figure}[t]
\centering
\includegraphics[width=3.5in]{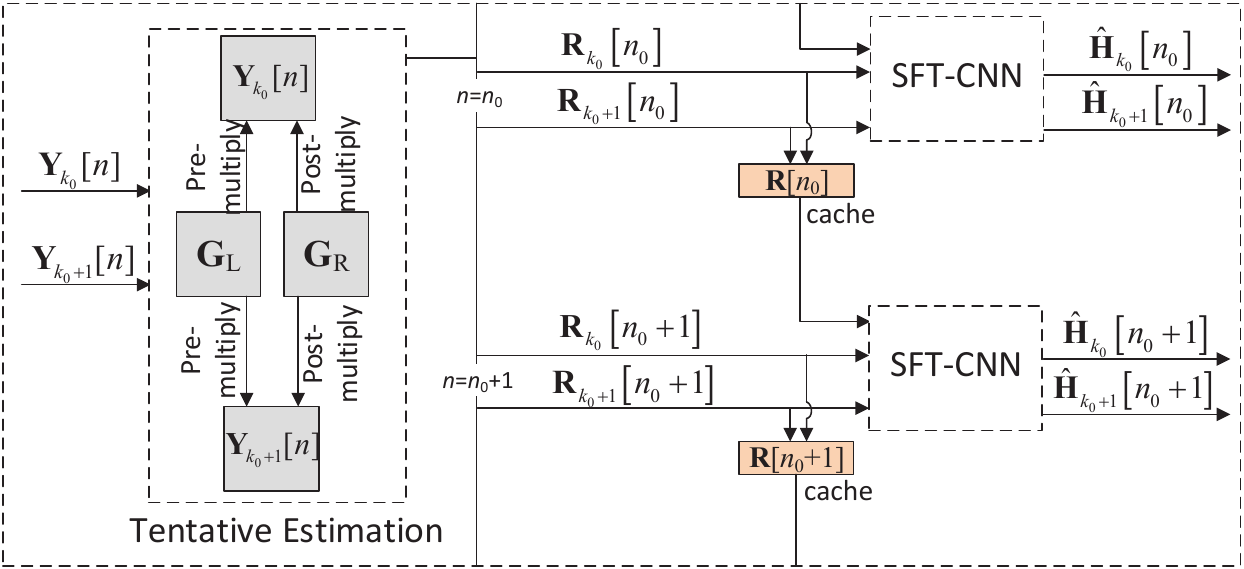}
\caption{SFT-CNN based channel estimation.}\label{SFT-CNN}
\end{figure}

In this section, we first develop a SFT-CNN based channel estimation approach, which further incorporates channel temporal correlation into the SF-CNN. Then the SFT-CNN is modified to the SPR-CNN to mitigate the huge spatial pilot overhead caused by large-scale antenna arrays.


For time-varying channels, the frequency-domain channel at the $k$th subcarrier in (\ref{eqn_H_k}) becomes \cite{R. W. Heath Jr.}
\setlength{\arraycolsep}{0.4em}
\begin{eqnarray}
\label{eqn_H_kt}
\mathbf{H}_k(t)=\sqrt{\frac{N_{\textrm{T}}N_{\textrm{R}}}{L}}\sum_{l=1}^L \alpha_{l}e^{-j2\pi(\tau_l f_{\textrm{s}}\frac{k}{K}-\nu_{l}t)}\mathbf{a}_{\textrm{R}}(\varphi_l)\mathbf{a}_{\textrm{T}}^{H}(\phi_l),
\end{eqnarray}
where $\nu_l$ denotes the Doppler shift of the $l$th path.

According to \cite{V. Va}, \cite{J. Choi}, the temporal correlation between channels in successive coherence intervals can be modeled as Gauss-Markov distribution
\setlength{\arraycolsep}{0.4em}
\begin{eqnarray}
\label{eqn_temp_corre}
\mathbf{H}_k[n]=\rho\mathbf{H}_k[n-1]+\sqrt{1-\rho^2}\boldsymbol{\Theta}[n], \quad n\in \mathbb{N}_{+}
\end{eqnarray}
where $\mathbf{H}_k[n]=\mathbf{H}_k(nT)$ is the discrete-time version of $\mathbf{H}_k(t)$ with $T$ denoting the length of the coherence interval, $0\leq\rho\leq 1$ denotes the temporal correlation coefficient, and $\boldsymbol{\Theta}[n]$ is a random matrix accounting for the innovation process with unit variance for each entry. (\ref{eqn_temp_corre}) clearly demonstrates that some inherence underlays the channel variation from the previous coherence interval to the current one and this correlation can be also exploited to improve the channel estimation accuracy in addition to the spatial and frequency correlation. In the following, we first elaborate the SFT-CNN based channel estimation.

\subsection{SFT-CNN based Channel Estimation}

As shown in Fig.~\ref{SFT-CNN}, we still consider channel estimation at $Q$ ($=2$) adjacent subcarriers, $k_{0}$ and $k_{0}+1$, for ease of illustration. In time-varying channels, the received pilots after combining at the receiver in (\ref{eqn_simple_Yk}) becomes
\setlength{\arraycolsep}{0.1em}
\begin{eqnarray}
\label{eqn_simple_Ykn}
\mathbf{Y}_k[n]\!=\!\sqrt{P}\mathbf{W}^H\mathbf{H}_k[n]\mathbf{F}\!+\!\tilde{\mathbf{N}}_k[n],\quad k\in\left\{k_0,k_0\!+\!1\right\}\!.
\end{eqnarray}

Similar to SF-CNN based channel estimation, $\mathbf{Y}_k[n]$ is then processed by the TE module, which generates the tentatively estimated channel matrices sequentially as
\setlength{\arraycolsep}{0.0em}
\begin{eqnarray}
\label{eqn_Rkn}
\mathbf{R}_k[n]=\sqrt{P}\mathbf{G}_{\textrm{L}}\mathbf{W}^H\mathbf{H}_k[n]\mathbf{F}\mathbf{G}_{\textrm{R}} +\mathbf{G}_{\textrm{L}}\tilde{\mathbf{N}}_k[n]\mathbf{G}_{\textrm{R}}.
\end{eqnarray}

Then a SFT-CNN further refines these tentatively estimated channel matrices by exploiting the spatial, frequency, and temporal correlation of channels simultaneously. As shown in Fig.~\ref{SFT-CNN}, we capture $S$ ($=2$) successive coherence intervals, $n_{0}$ and $(n_{0}+1)$, to describe the channel estimation procedure. In the $n_{0}$th coherence interval, the tentatively estimated channel matrices, $\mathbf{R}_{k_{0}}[n_{0}]$ and $\mathbf{R}_{k_{0}+1}[n_{0}]$, are input into the SFT-CNN. A copy of $\mathbf{R}_{k_{0}}[n_{0}]$ and $\mathbf{R}_{k_{0}+1}[n_{0}]$ is stored in the cache in order to be used in the next coherence interval. In $(n_{0}+1)$th coherence interval, the SFT-CNN receives tentatively estimated channel matrices, $\mathbf{R}_{k_{0}}[n_{0}+1]$ and $\mathbf{R}_{k_{0}+1}[n_{0}+1]$, as well as fetches $\mathbf{R}_{k_{0}}[n_{0}]$ and $\mathbf{R}_{k_{0}+1}[n_{0}]$ from the cache to perform joint processing and obtain the estimated channel matrices as
\setlength{\arraycolsep}{0.0em}
\begin{eqnarray}
\label{eqn_SFTCNNmap}
&&\left\{\hat{\mathbf{H}}_{k_{0}}[n_{0}+1], \hat{\mathbf{H}}_{k_{0}+1}[n_{0}+1]\right\}\nonumber\\ &&=\!f_{\Psi}\!\left(\mathbf{R}_{k_{0}}[n_{0}],\mathbf{R}_{k_{0}\!+\!1}[n_{0}],\mathbf{R}_{k_{0}}[n_{0}\!\!+\!\!1],\mathbf{R}_{k_{0}\!+\!1}[n_{0}\!\!+\!\!1];\Psi\right)\!,\quad
\end{eqnarray}
where $\Psi$ denotes the parameter set of the SFT-CNN. Meanwhile, the cache is updated by replacing $\mathbf{R}_{k_{0}}[n_{0}]$ and $\mathbf{R}_{k_{0}+1}[n_{0}]$ with $\mathbf{R}_{k_{0}}[n_{0}+1]$ and $\mathbf{R}_{k_{0}+1}[n_{0}+1]$. In each coherence interval, the same SFT-CNN is used since it has learned the general channel temporal correlation instead of the specific relationship between channels in two successive coherence intervals.

After summarizing the channel estimation procedure, we focus on the offline training of the SFT-CNN. Similar to SF-CNN, the training set consisting of $N_{\textrm{tr}}$ samples is generated according to certain channel model in the simulation environment with $\left(\underline{\mathbf{R}}_i,\underline{\mathbf{H}}_i\right)$ denoting the $i$th sample. $\underline{\mathbf{R}}_i\in\mathbb{C}^{N_{\textrm{R}}\times N_{\textrm{T}}\times 4}$ is a three-dimensional matrix composed of the tentatively estimated channel matrices in the $n_{0}'$th and $(n_{0}'+1)$th coherence intervals collected through (\ref{eqn_Rkn}), that is $\underline{\mathbf{R}}_i=\left[\mathbf{R}_{k_{0}'}^{i}[n_{0}'],\mathbf{R}_{k_{0}'+1}^{i}[n_{0}'],\mathbf{R}_{k_{0}'}^{i}[n_{0}'+1],\mathbf{R}_{k_{0}'+1}^{i}[n_{0}'+1]\right]$ with $n_{0}'\in\mathbb{N}_{+}$. $\underline{\mathbf{H}}_i\in\mathbb{C}^{N_{\textrm{R}}\times N_{\textrm{T}}\times 2}$ is also a three-dimensional matrix composed of the scaled true channel matrices in the $(n_{0}'+1)$th coherence interval, that is $\underline{\mathbf{H}}_i=\left[\frac{\mathbf{H}_{k_{0}'}^{i}[n_{0}'+1]}{c},\frac{\mathbf{H}_{k_{0}'+1}^{i}[n_{0}'+1]}{c}\right]$. As before, $c>0$ is the scaling constant to make the value range of the real and imaginary parts of all the target data, $\underline{\mathbf{H}}_i$, match the activation function applied in the output layer of the SFT-CNN. Then $\underline{\mathbf{R}}_i$ is fed into the SFT-CNN to approximate the corresponding scaled true channels $\underline{\mathbf{H}}_i$. The architecture of the SFT-CNN is similar to the SF-CNN except that it has the additional input from the previous coherence interval. With the estimated scaled channel matrices, $\frac{\hat{\mathbf{H}}_{k_{0}'}^{i}[n_{0}'+1]}{c},\frac{\hat{\mathbf{H}}_{k_{0}'+1}^{i}[n_{0}'+1]}{c}$, the objective of the SFT-CNN offline training is to minimize the MSE loss function
\setlength{\arraycolsep}{0.0em}
\begin{eqnarray}
\label{eqn_mse2}
\textrm{MSE}_{\textrm{Loss}}&&=\frac{1}{N_{\textrm{tr}}c^2}\sum_{i=1}^{N_{\textrm{tr}}}\sum_{q=1}^{2}\bigl\|\mathbf{H}_{k_{0}'+q-1}^{i}[n_{0}'+1] \nonumber\\ &&\quad-\hat{\mathbf{H}}_{k_{0}'+q-1}^{i}[n_{0}'+1]\bigr\|_{F}^2.
\end{eqnarray}

Compared to SF-CNN, SFT-CNN only increases the computational complexity for the first convolutional layer by $S$ times, which is a quite minor part in the total computational complexity according to (\ref{eqn_complexity_CNN}) and Table~\ref{SFCNN_para}. In contrast, if further incorporating temporal correlation, the complexity of MMSE channel estimation in (\ref{eqn_complexity_MMSE}) becomes
\setlength{\arraycolsep}{0.0em}
\begin{eqnarray}
\label{eqn_complexity_MMSE2}
C_{\textrm{MMSE}}\sim\mathcal{O}(S^3Q^3N_{\textrm{T}}^3N_{\textrm{R}}^3),
\end{eqnarray}
which will be increased significantly even with $S=2$. Therefore, SFT-CNN provides a simple and efficient way to utilize the channel spatial, frequency, and temporal correlation simultaneously to improve the channel estimation accuracy.

\subsection{SFT-CNN or LSTM-CNN?}

The LSTM-CNN would naturally come to mind when we consider temporal correlation. In this subsection, we will compare SFT-CNN and LSTM-CNN in terms of estimation accuracy and complexity, which explains why the SFT-CNN in used instead of LSTM-CNN.

\emph{1) Architecture: }The LSTM-CNN consists of one input layer, three convolutional LSTM layers, seven convolutional layers, and one output layer, where the processing of each convolutional LSTM layer follows the operation in \cite{X. Shi}. The detailed architecture of the LSTM-CNN is listed in Table~\ref{table_LSTMCNN}.

\begin{table}[!t]
\centering
\caption{Architecture of the LSTM-CNN}
\label{table_LSTMCNN}
\begin{tabular}{c|c|c|c|c|c}
\hline
 \makecell{Layer} & \makecell{Type} & \makecell{Number\\of filters} & \makecell{Size\\ of filters} & \makecell{Activation\\ function} & \makecell{Recurrent\\ activation\\ function}\\
\hline
 1 & Input & - & - & - & - \\
\hline
 2$\sim$4 & Conv. LSTM & $4$ & $3\times3$ & tanh & tanh \\
\hline
 5$\sim$11 & Conv. & $64$ & $3\times3$ & ReLU & - \\
\hline
 12 & Output & $2Q$ & $3\times3$ & tanh & - \\
\hline
\end{tabular}
\end{table}

\begin{figure}[!t]
\centering
\includegraphics[trim=0 0 0 0, width=3.6in]{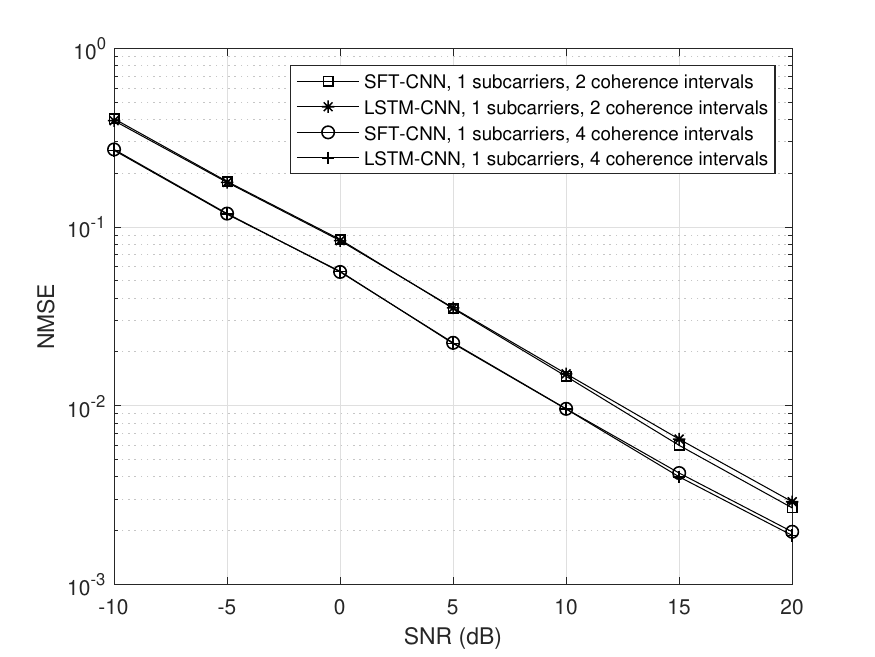}
\caption{NMSE for SFT-CNN and LSTM-CNN based channel estimation.}\label{nmse_SFT_LSTM}
\end{figure}

\emph{2) Channel Estimation Accuracy: }To measure the channel estimation performance, we use the normalized MSE (NMSE), defined as,
\begin{equation}
\begin{aligned}
\label{eqn_nmse}
\textrm{NMSE}=\mathbb{E}_{\mathbf{H}}\left\{\|\mathbf{H}-\hat{\mathbf{H}}\|_{F}^2/\|\mathbf{H}\|_{F}^2\right\},
\end{aligned}
\end{equation}
where $\mathbf{H}$ and $\hat{\mathbf{H}}$ refer to the true and estimated channels, respectively. Fig.~\ref{nmse_SFT_LSTM} plots the NMSE performance of SFT-CNN and LSTM-CNN based channel estimation, where two and four coherence intervals are involved, respectively. From this figure, SFT-CNN and LSTM-CNN achieve almost the same performance in the whole signal-to-noise ratio (SNR) regime.

\begin{table}[!t]
\centering
\caption{Time Complexity of Training and Testing}
\label{table_timecomplexity}
\begin{tabular}{p{1.5cm}<{\centering}|p{1cm}<{\centering}|p{1cm}<{\centering}|p{1.5cm}<{\centering}|p{1.5cm}<{\centering}}
\hline
\multirow{2}{*}{} & \multicolumn{2}{c|}{\makecell{Training time\\ seconds/epoch}} & \multicolumn{2}{c}{\makecell{Testing time\\ seconds/channel realization}} \\
\cline{2-5}
 ~ & $N_c=2$ & $N_c=4$ & $N_c=2$ & $N_c=4$ \\
\hline
 SFT-CNN & $22$ & $23$  & $1.51\times10^{-4}$ & $1.7\times10^{-4}$  \\
\hline
 LSTM-CNN & $34$ & $48$ & $3.53\times10^{-4}$ & $5.14\times10^{-4}$  \\
\hline
\end{tabular}
\end{table}

\emph{3) Training and Testing Complexity: }With the same NMSE performance, the complexity becomes an important metric for method selection. Table~\ref{table_timecomplexity} provides the time complexity of training and testing for SFT-CNN and LSTM-CNN on the NVIDIA GeForce GTX 1080 Ti GPU, where $N_c$ denotes the number of coherence intervals involved in the CNNs. It is clear that LSTM-CNN consumes much more time than SFT-CNN in both training and testing stages.

Based on the above discussion, the LSTM-CNN has no performance advantages in exploiting channel temporal correlation but requires much higher complexity for training and testing. Since the SFT-CNN can also capture the channel temporal correlation efficiently with relatively low complexity, this simple pure CNN architecture is preferable for the channel estimation task in this paper.

\subsection{SPR-CNN based Channel Estimation}

\begin{figure}[t]
\centering
\includegraphics[width=3.5in]{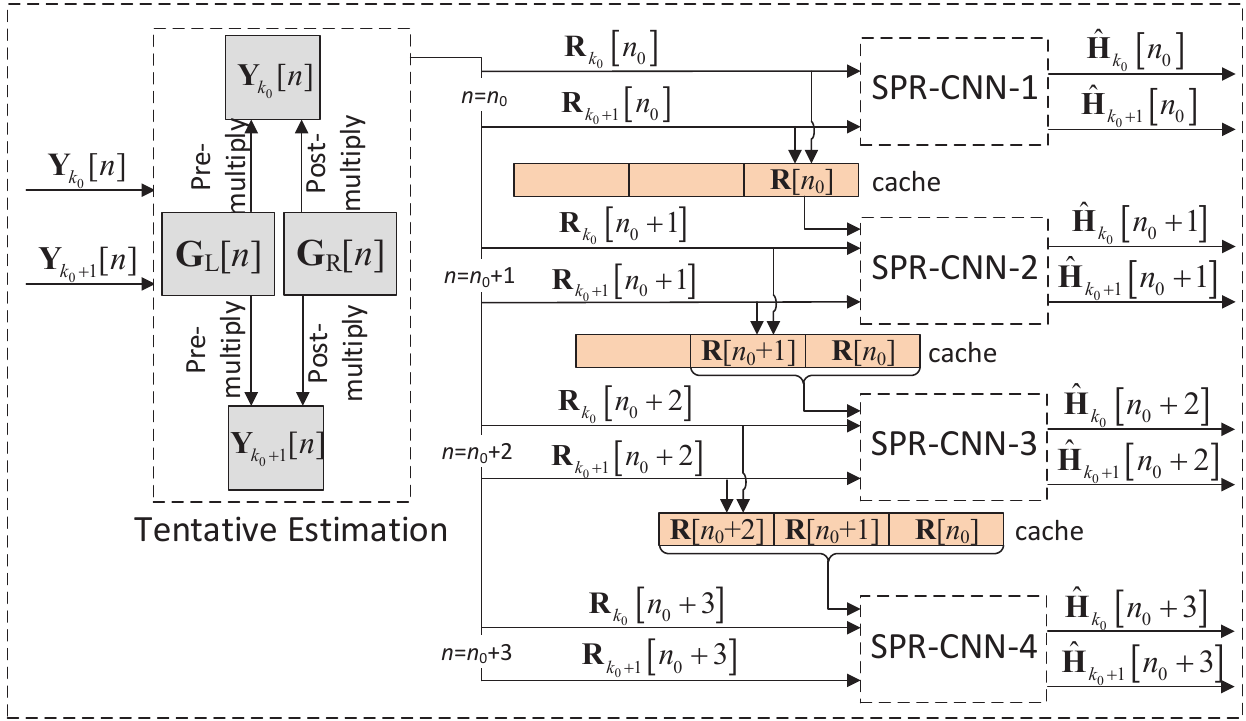}
\caption{SPR-CNN based channel estimation.}\label{SPR-CNN}
\end{figure}

Large-scale array antennas at both the transmitter and the receiver incur huge pilot overhead in spatial domain. In this subsection, we design the SPR-CNN based channel estimation, which uses much fewer pilots but still guarantees the fairly good accuracy.

The basic idea of the SPR-CNN based channel estimation can be summarized as follows:
\begin{itemize}[\IEEEsetlabelwidth{Z}]
\item[1)] Group $D$ successive coherence intervals as a CEU, in which channel correlation is utilized to reduce the spatial pilot overhead. Different CEUs are non-overlapped.

\item[2)] In the first coherence interval of each CEU, use full spatial pilot overhead for channel estimation.\footnote{Full spatial pilot overhead means that the number of beamforming vectors is equal to the number of transmit antennas and the number of combining vectors is equal to the number of receive antennas.} Then the pilot overhead is reduced in the subsequent coherence intervals.

\item[3)] For the first coherence interval, the receiver uses the currently received pilots to estimate the current channels. For the rest of coherence intervals, the receiver uses the currently and all previously received pilots in this CEU to jointly estimate the current channels.
\end{itemize}

Here is the detailed channel estimation procedure. Different beamforming and combining matrices are employed in different coherence intervals of each CEU. As shown in Fig.~\ref{SPR-CNN}, the received pilots after combining at the receiver in (\ref{eqn_simple_Ykn}) becomes
\setlength{\arraycolsep}{0.0em}
\begin{eqnarray}
\label{eqn_simple_Ykn_differ}
\mathbf{Y}_k[n]=\sqrt{P}\mathbf{W}^H[n]\mathbf{H}_k[n]\mathbf{F}[n]+\tilde{\mathbf{N}}_k[n],
\end{eqnarray}
where $\mathbf{F}[n]\in\mathbb{C}^{N_{\textrm{T}}\times M_{\textrm{T}}[n]}$ and $\mathbf{W}[n]\in\mathbb{C}^{N_{\textrm{R}}\times M_{\textrm{R}}[n]}$ denote the beamforming matrix and combining matrix, respectively, in the $n$th coherence interval. The corresponding spatial pilot overhead is given by
\setlength{\arraycolsep}{0.0em}
\begin{eqnarray}
\label{eqn_pilot_n}
p[n]=M_{\textrm{T}}[n]\left\lceil\frac{M_{\textrm{R}}[n]}{N_{\textrm{R}}^{\textrm{RF}}}\right\rceil.
\end{eqnarray}
From (\ref{eqn_pilot_n}), the spatial pilot overhead can be saved by reducing $M_{\textrm{T}}[n]$ or/and $M_{\textrm{R}}[n]$.

$\mathbf{Y}_k[n]$ is first processed by the TE module and the tentatively estimated channel matrix is given by
\setlength{\arraycolsep}{0.0em}
\begin{eqnarray}
\label{eqn_Rkn_differ}
\mathbf{R}_k[n]&&=\mathbf{G}_{\textrm{L}}[n]\mathbf{Y}_k[n]\mathbf{G}_{\textrm{R}}[n]\nonumber\\
&&=\sqrt{P}\mathbf{G}_{\textrm{L}}[n]\mathbf{W}^H[n]\mathbf{H}_k[n]\mathbf{F}[n]\mathbf{G}_{\textrm{R}}[n] \nonumber\\ &&\quad+\mathbf{G}_{\textrm{L}}[n]\tilde{\mathbf{N}}_k[n]\mathbf{G}_{\textrm{R}}[n],
\end{eqnarray}
where $\mathbf{G}_{\textrm{L}}[n]$ and $\mathbf{G}_{\textrm{R}}[n]$ are also changed along with the coherence interval index and are expressed as
\setlength{\arraycolsep}{0.0em}
\begin{eqnarray}
\label{eqn_GLn}
\mathbf{G}_{\textrm{L}}[n]=
\begin{cases}
\mathbf{W}[n], & M_{\textrm{R}}[n]<N_{\textrm{R}},\\
(\mathbf{W}[n]\mathbf{W}^H[n])^{-1}\mathbf{W}[n], & M_{\textrm{R}}[n]\geq N_{\textrm{R}},
\end{cases}
\end{eqnarray}
and
\setlength{\arraycolsep}{0.0em}
\begin{eqnarray}
\label{eqn_GRn}
\mathbf{G}_{\textrm{R}}[n]=
\begin{cases}
\mathbf{F}^H[n], & M_{\textrm{T}}[n]<N_{\textrm{T}},\\
\mathbf{F}^H[n](\mathbf{F}[n]\mathbf{F}^H[n])^{-1}, & M_{\textrm{T}}[n]\geq N_{\textrm{T}}.
\end{cases}
\end{eqnarray}

Then the $\mathbf{R}_k[n]$ is processed by the CNN estimation part. We consider that every $D$ ($=4$) successive coherence intervals are grouped as a CEU and capture a certain CEU with the $n_{0}$th to the $(n_{0}+3)$th coherence intervals, as shown in Fig.~\ref{SPR-CNN}. There are four SFT-CNNs with different input and output, called SPR-CNN-$1$, $2$, $3$, and $4$, respectively. In the $n_{0}$th coherence interval, full spatial pilot overhead, i.e., $M_{\textrm{T}}[n]=N_{\textrm{T}}$ and $M_{\textrm{R}}[n]=N_{\textrm{R}}$, is used to provide accurate channel information for all coherence intervals of this CEU. After the TE module, $\mathbf{R}_{k_0}[n_0]$ and $\mathbf{R}_{k_0+1}[n_0]$ are input into the SPR-CNN-$1$, which generates the finally estimated channel matrices, $\hat{\mathbf{H}}_{k_0}[n_0]$ and $\hat{\mathbf{H}}_{k_0+1}[n_0]$. Meanwhile, a copy of $\mathbf{R}_{k_{0}}[n_{0}]$ and $\mathbf{R}_{k_{0}+1}[n_{0}]$ is stored in the cache to provide additional channel information for the channel estimation of subsequent coherence intervals. In the $(n_{0}+1)$th coherence interval, the dimensions of $\mathbf{F}[n]$ and $\mathbf{W}[n]$ will be reduced to save the pilot overhead, i.e., $M_{\textrm{T}}[n]<N_{\textrm{T}}$ and $M_{\textrm{R}}[n]<N_{\textrm{R}}.$\footnote{Different pilot overheads may be used in different coherence intervals of a CEU. But for each coherence interval, it and its counterparts in other CEUs should use the same pilot overhead.} $\mathbf{R}_{k_0}[n_0]$ and $\mathbf{R}_{k_0+1}[n_0]$ stored in the cache along with $\mathbf{R}_{k_0}[n_0+1]$ and $\mathbf{R}_{k_0+1}[n_0+1]$ are simultaneously input into SPR-CNN-$2$ to obtain $\hat{\mathbf{H}}_{k_0}[n_0+1]$ and $\hat{\mathbf{H}}_{k_0+1}[n_0+1]$. A copy of $\mathbf{R}_{k_{0}}[n_{0}+1]$ and $\mathbf{R}_{k_{0}+1}[n_{0}+1]$ is also stored in the cache in addition to $\mathbf{R}_{k_{0}}[n_{0}]$ and $\mathbf{R}_{k_{0}+1}[n_{0}]$. All matrices stored in the cache are used for the joint channel estimation of the $(n_{0}+2)$th coherence interval. Channel estimations in the $(n_{0}+2)$th and $(n_{0}+3)$th coherence intervals are similar to that in the $(n_{0}+1)$th coherence interval. After channel estimation in the $(n_{0}+3)$th coherence interval, the cache will be emptied and then used for the next CEU. From Fig.~\ref{SPR-CNN}, four different SPR-CNNs are employed for respective coherence intervals in a CEU and reused for all CEUs. The architecture and training process of the SPR-CNNs are similar to SFT-CNN except that the numbers of input matrices are different for different SPR-CNNs. An intuitional description of the SPR-CNN based channel estimation is given by Algorithm 1.

\begin{algorithm}[h]
\label{alg:SPR_CNN}
\caption{SPR-CNN based Channel Estimation}
{\bf Input:}
The total number of CEUs $M_{\textrm{CEU}}$, the number of coherence intervals in each CEU $D$, spatial pilot overhead from the second to $D$th coherence intervals of each CEU\\
{\bf Output:}
Estimated channel matrices\\
{\bf Procedure:}
\begin{algorithmic}[1]
\STATE Initialize the CEU and coherence interval indices as $m=1$ and $d=1$;
\STATE Train $D$ different SPR-CNNs for the first to $D$th coherence intervals;
\FOR{$m\in[1,M_{\textrm{CEU}}]$}
\FOR{$d\in[1,D]$}
\IF{$d=1$}
\STATE Use full spatial pilot overhead;
\STATE Tentatively estimate channels according to (\ref{eqn_Rkn_differ});
\STATE Store the tentatively estimated channel matrices in the cache;
\STATE Input the tentatively estimated channel matrices of the first coherence interval into SPR-CNN-$1$ to obtain the estimated channel matrices of the first coherence interval;
\ELSE
\STATE Use reduced spatial pilot overhead;
\STATE Tentatively estimate channels according to (\ref{eqn_Rkn_differ});
\STATE Store the tentatively estimated channel matrices in the cache (Invalid for $d=D$);
\STATE Input the tentatively estimated channel matrices from the first to $d$th coherence intervals into SPR-CNN-$d$ to obtain the estimated channel matrices of the $d$th coherence interval;
\ENDIF
\ENDFOR
\STATE Empty the cache and reset $d=1$
\ENDFOR
\RETURN the estimated channel matrices of $M_{\textrm{CEU}}$ CEUs
\end{algorithmic}
\end{algorithm}

Among the four SPR-CNNs, SPR-CNN-$4$ has the highest complexity with the most input matrices. But it just increases the complexity of the first convolutional layer by $D$ ($=4$) times compared to the SF-CNN in Section III, which causes limited impact on the total computational complexity. Therefore, SPR-CNN based channel estimation saves the spatial pilot overhead effectively while only increases the complexity moderately.

We design SPR-CNN based channel estimation aiming to reduce the pilot overhead in spatial domain significantly while still guarantee fairly good channel estimation accuracy. In each CEU with $D$ coherence intervals, full pilot overhead is used in the first coherence interval while reduced pilot overhead is used in the remaining coherence intervals. The first coherence interval provides complete channel information that is quite helpful to channel estimation of all coherence intervals in this CEU. It is clear that the average pilot overhead of a CEU reduces as $D$ increases. However, increasing $D$ will also degrade the average channel estimation accuracy since the effect of the complete channel information provided by the first coherence interval weakens along with the evanescent temporal correlation. Therefore, our architecture of SPR-CNN includes four coherence intervals in a CEU to achieve a good tradeoff between the pilot overhead and the estimation accuracy. In addition, if using the LSTM-CNN to perform the channel estimation in Fig.~\ref{SPR-CNN}, four LSTM-CNNs with the architecture listed in Table~\ref{table_LSTMCNN} are needed. As discussed in Section IV.B, the LSTM-CNN has no performance superiority over the simple SFT-CNN architecture but causes much higher complexity and thus is not suitable for the SPR-CNN based channel estimation.

\section{Numerical Results}

In this section, we present simulation results of the proposed CNN based channel estimation approaches and compare them with non-ideal MMSE using the estimated covariance matrix and ideal MMSE using the true covariance matrix. We set the number of antennas at the transmitter, $N_{\textrm{T}}=32$, the number of antennas at the receiver, $N_{\textrm{R}}=16$, and the numbers of RF chains at the transmitter and the receiver $N_{\textrm{T}}^{\textrm{RF}}=N_{\textrm{R}}^{\textrm{RF}}=2$. $\mathbf{F}$(or $\mathbf{F}[n]$) and $\mathbf{W}$(or $\mathbf{W}[n]$) are set as the first $M_{\textrm{T}}$ (or $M_{\textrm{T}}[n]$) columns of an $N_{\textrm{T}}\times N_{\textrm{T}}$ discrete Fourier transform (DFT) matrix and the first $M_{\textrm{R}}$ (or $M_{\textrm{R}}[n]$) columns of an $N_{\textrm{R}}\times N_{\textrm{R}}$ DFT matrix. In Section V.A and Section V.B, we set $M_{\textrm{T}}=32$ and $M_{\textrm{R}}=16$. The settings of $M_{\textrm{T}}[n]$ and $M_{\textrm{R}}[n]$ will be introduced in Section V.C.

The channel data are generated according to the 3rd Generation Partnership Project (3GPP) TR 38.901 Release 15 channel model \cite{3GPP}. Specifically, we use the clustered delay line models with the carrier frequency, $f_c=28$ GHz, the sampling rate, $f_s=100$ MHz, the number of main paths, $L=3$, and the number of subcarriers, $K=64$.

For the SF-CNN, the training set, validation set, and testing set contain $81$,$000$, $9$,$000$, and $19$,$000$ samples, respectively. The parameters of each layer are set as Table~\ref{table_SFCNN}. Adam is used as the optimizer. The epochs are set as $800$ while the corresponding learning rates are $10^{-4}$ for the first $200$ epochs, $5\times10^{-5}$ for the next $400$ epochs, and $10^{-5}$ for the last $200$ epochs, respectively.\footnote{Although optimizer Adam can adaptively adjust the learning rate for each epoch by considering the momentum as well as the gradients, we still manually set the learning rate at the specific epochs to speed up the convergence of offline training.} The batch size is $128$. The scaling constant is set as $c=2$. The SFT-CNN and SPR-CNN use the same parameters as the SF-CNN except that the numbers of input matrices are different.

\subsection{SF-CNN based Channel Estimation}

\begin{figure}[!t]
\centering
\includegraphics[trim=15 15 0 10, width=3.6in]{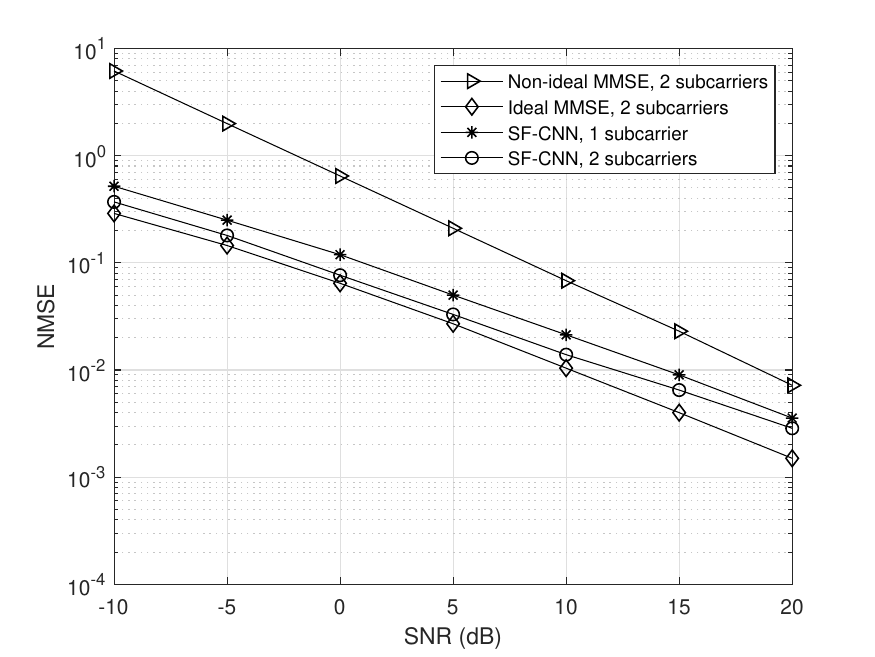}
\caption{NMSE versus SNR for the SF-CNN based channel estimation and MMSE channel estimation.}\label{nmse_SF_CNN}
\end{figure}

Fig.~\ref{nmse_SF_CNN} shows the NMSE performance versus SNR of the proposed SF-CNN based channel estimation and MMSE channel estimation over two adjacent subcarriers in the urban micro (UMi) street non-line of sight (NLOS) scenario.\footnote{We train different neural networks separately for different SNRs based on the same channel dataset. Specifically, we first generate a channel dataset according to the 3GPP TR 38.901 Release 15 channel model. Then we use these channel data to generate the received pilot signals in different SNRs, which are used to train the neural networks for corresponding SNRs. The testing set is generated similarly.} The performance of the SF-CNN based approach at single subcarrier is also plotted to demonstrate that frequency correlation is helpful to improve the channel estimation accuracy. Through offline training, the SF-CNN based channel estimation outperforms the non-ideal MMSE with estimated covariance matrix significantly yet requiring lower estimation complexity according to this figure and Section III.B. Moreover, the performance of the SF-CNN based approach is very close to the ideal MMSE with true covariance matrix, especially at the low and medium SNRs.

\begin{figure}[!t]
\centering
\includegraphics[trim=15 15 0 10, width=3.6in]{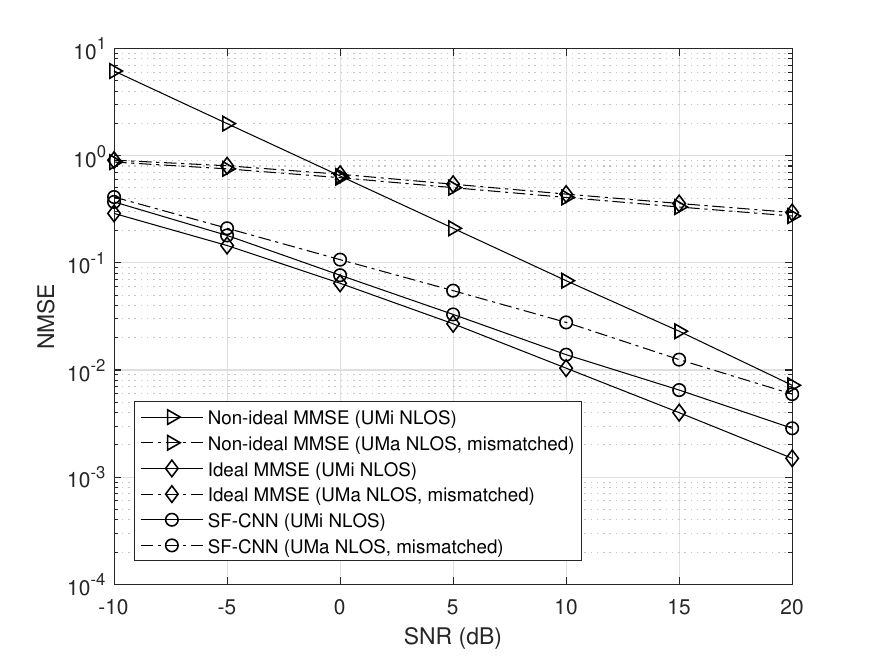}
\caption{Robustness of MMSE and SF-CNN based approaches to different scenarios.}\label{SF_CNN_robustness}
\end{figure}

The robustness of the MMSE and proposed SF-CNN based approaches is shown in Fig.~\ref{SF_CNN_robustness}. The joint channel estimation over two subcarriers is considered. The SF-CNN is trained in the UMi street NLOS scenario and is tested in both UMi street NLOS scenario and urban macro (UMa) NLOS scenario. For the MMSE, its covariance matrix is calculated in the UMi street NLOS scenario and then the channel matrix is estimated in both UMi street NLOS scenario and UMa NLOS scenario. According to \cite{3GPP}, the power, delay, and angle profiles of UMa NLOS scenario are quite different from those of UMi street NLOS scenario. The channel statistics are unknown to both SF-CNN and MMSE when they predict the channels in the UMa NLOS scenario. From this figure, the SF-CNN based channel estimation exhibits good robustness when facing the significantly different channel statistics. Even under the mismatched UMa NLOS scenario, the SF-CNN based approach still outperforms the non-ideal MMSE without mismatch. In contrast, due to the high dependence on channel statistics, both the ideal and non-ideal MMSE fail to cope with the change of channel statistics and suffer significant performance loss.

\begin{figure}[!t]
\centering
\includegraphics[trim=15 15 0 10, width=3.6in]{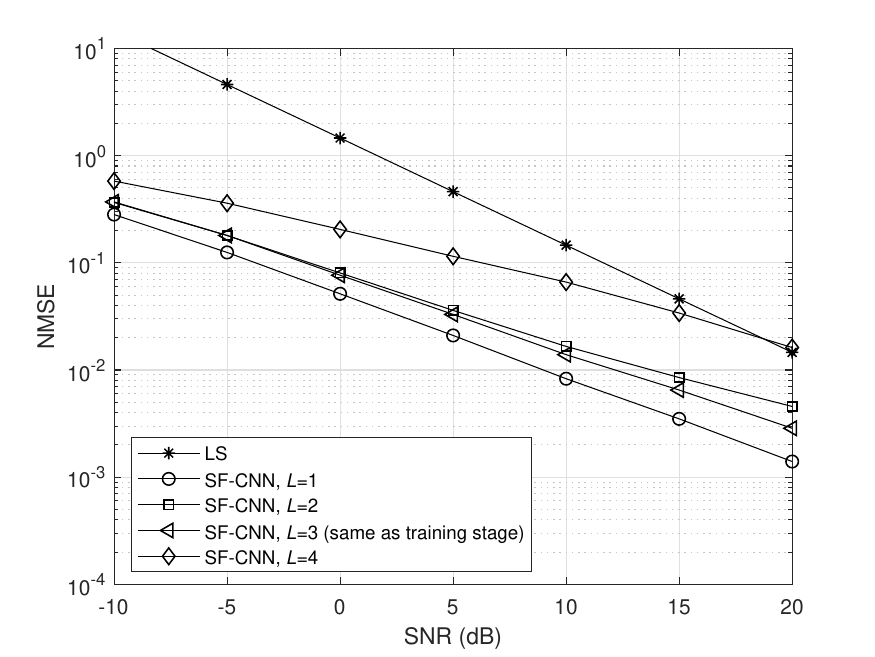}
\caption{Robustness of SF-CNN based approach to different numbers of main paths.}\label{SF_CNN_robustness_diffpath}
\end{figure}

The number of main paths, $L$, may change in the real scenario. Here, we test the robustness of the SF-CNN based approach to different numbers of main paths, which is plotted in Fig.~\ref{SF_CNN_robustness_diffpath} with $L=3$ in the training stage and $L=1,2,3,4$ in the testing stage. The case of $L=3$ in the testing stage is matched with the training stage and thus acts as the baseline for comparison. LS channel estimation is also plotted as an upper bound. The SF-CNN based approach with $L=1$ performs better than the baseline, which is because the channel structure becomes quite simple and thus facilitates the CNN denoising. When $L$ is increased to $2$, the performance is slightly degraded compared to the baseline. Although there is a significant performance degeneration when $L$ becomes $4$, the proposed approach still outperforms LS channel estimation when SNR$<18$ dB. In fact, $L=1$ and $2$ can be regarded as two special cases of $L=3$. The offline trained SF-CNN with $L=3$ is able to identify the channels with $L=1$ and $2$ and thus exhibits good robustness. In contrast, the case of $L=4$ leads to a significant change of channel structure, to which the offline trained model is less robust.

\subsection{SFT-CNN based Channel Estimation}

\begin{figure}[!t]
\centering
\includegraphics[trim=15 15 0 10, width=3.6in]{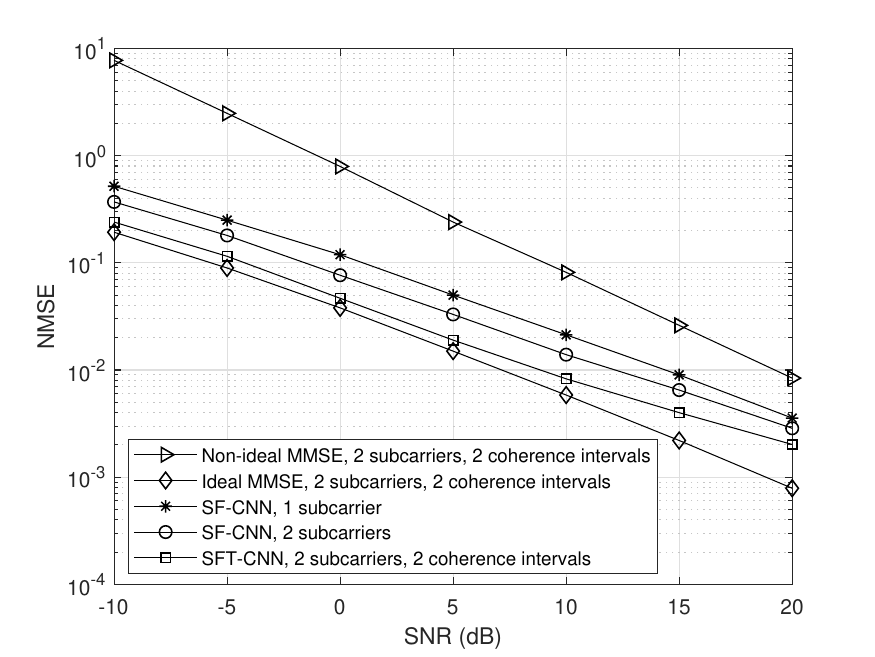}
\caption{NMSE versus SNR for the SFT-CNN based channel estimation and MMSE channel estimation.}\label{nmse_SFT_CNN}
\end{figure}

Fig.~\ref{nmse_SFT_CNN} shows the NMSE performance versus SNR of the SF-CNN, SFT-CNN, and MMSE based channel estimation approaches in the UMi street NLOS scenario. The MMSE and proposed SFT-CNN based approaches utilize the channel information of the previous coherence interval to jointly estimate the current channels over two subcarriers while the SF-CNN based approach does not incorporate temporal correlation. By comparing the circle and square curves, we can clearly see the effect of temporal correlation on improving the NMSE performance with the SFT-CNN based channel estimation. The comparison between the star and square curves demonstrates that just utilizing the additional channel information of adjacent subcarrier and coherence interval can improve the estimation accuracy remarkably while only increases the complexity slightly. With the same channel correlation information, the SFT-CNN based approach outperforms the non-ideal MMSE significantly in both performance and complexity at the cost of offline training. It is even very close to the ideal MMSE at the low and medium SNRs.

\begin{figure}[!t]
\centering
\includegraphics[trim=15 15 0 10, width=3.6in]{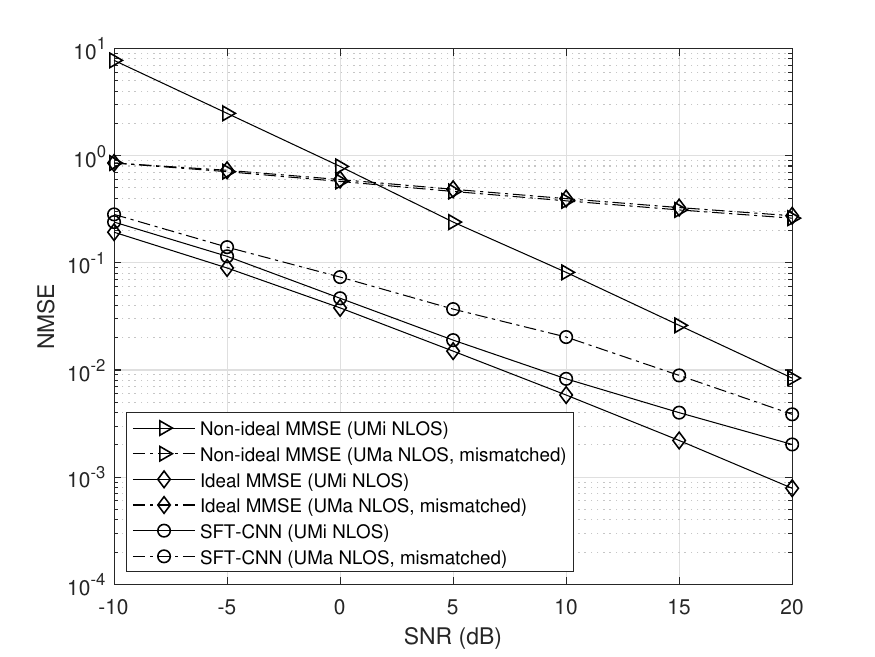}
\caption{Robustness of MMSE and SFT-CNN based approaches to different scenarios.}\label{SFT_CNN_robustness}
\end{figure}

In Fig.~\ref{SFT_CNN_robustness}, we demonstrate the robustness of the MMSE and proposed SFT-CNN based approaches using the channel correlation of two subcarriers and two coherence intervals. Similar to Fig.~\ref{SF_CNN_robustness}, both the covariance matrix calculation of the MMSE and the offline training of the SFT-CNN are performed in the UMi street NLOS scenario with the maximum Doppler spread $f_{d}=1,400$ Hz. When the testing scenario is unchanged, all these approaches can achieve good performance. If we test them in a mismatched scenario, i.e., UMa NLOS scenario with $f_{d}=1,800$ Hz in Fig.~\ref{SFT_CNN_robustness}, the performance of both ideal and non-ideal MMSE severely degrades since the mismatched covariance matrix is counterproductive when refining the LS estimated channels. In contrast, the SFT-CNN has learned the more inherent channel structure and thus exhibits superior robustness to the significantly different scenarios.

\begin{figure}[!t]
\centering
\includegraphics[trim=15 15 0 10, width=3.6in]{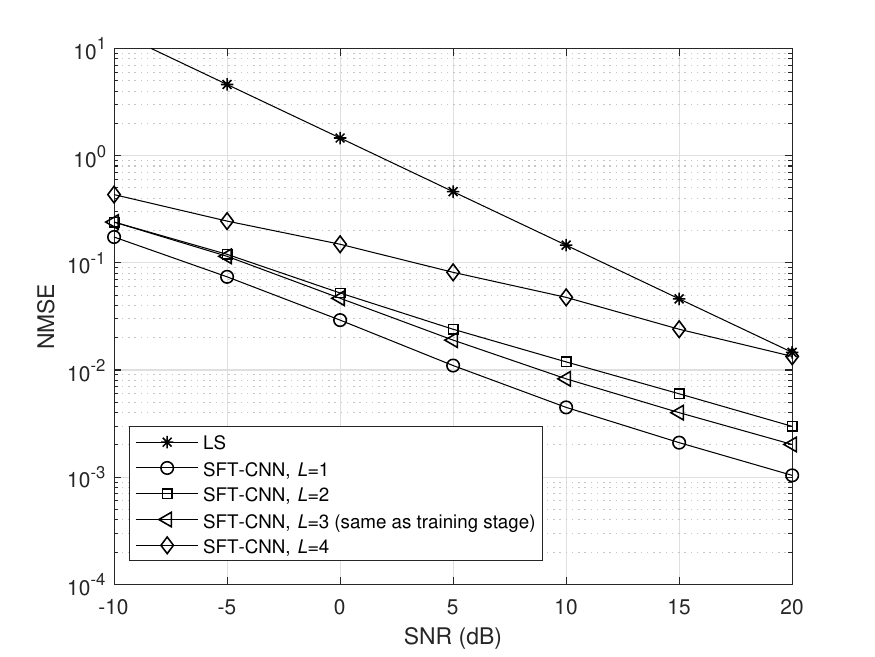}
\caption{Robustness of SFT-CNN based approach to different numbers of main paths.}\label{SFT_CNN_robustness_diffpath}
\end{figure}

The robustness of SFT-CNN based approach to different numbers of main paths is evaluated in Fig.~\ref{SFT_CNN_robustness_diffpath}, which exhibits similar results to Fig.~\ref{SF_CNN_robustness_diffpath}. Nevertheless, temporal correlation is still helpful to improve the channel estimation accuracy even with mismatched number of main paths by comparing Fig.~\ref{SF_CNN_robustness_diffpath} and Fig.~\ref{SFT_CNN_robustness_diffpath}.\footnote{Robustness is an important issue since a robust DNN can avoid the complex online fine-tuning process. There are infinite channel statistics or scenarios in reality and a DNN robust to as many cases as possible is expected to design. This is a challenging task and needs more research attention since the settings of the NN architecture and various hyper-parameters are still unclear for the complicated mmWave MIMO channels.}

\subsection{SPR-CNN based Channel Estimation}

\begin{figure}[!t]
\centering
\includegraphics[trim=15 15 0 10, width=3.6in]{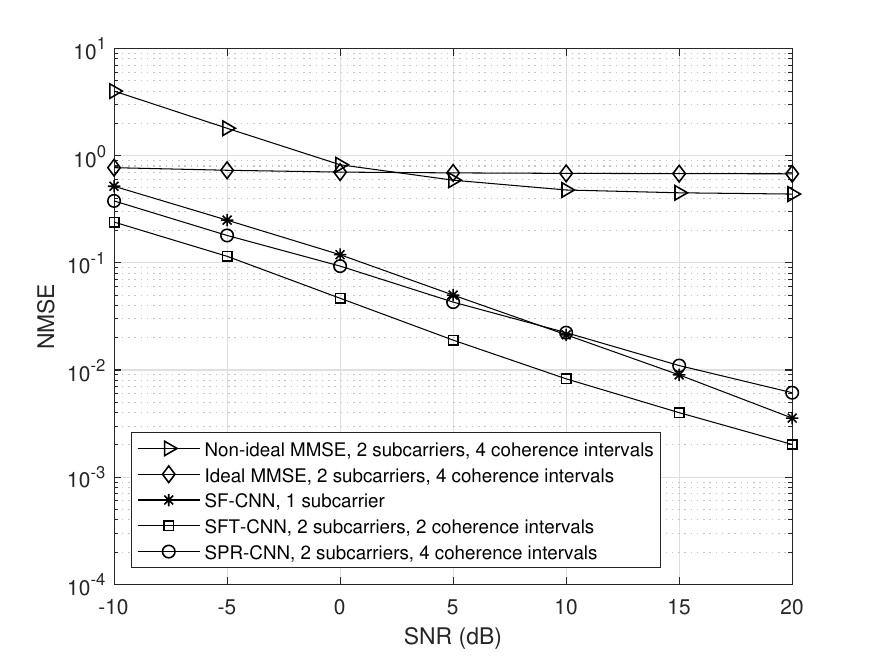}
\caption{NMSE versus SNR for the SPR-CNN based channel estimation and MMSE channel estimation.}\label{nmse_SPR_CNN}
\end{figure}

In Fig.~\ref{nmse_SPR_CNN}, we verify the effectiveness of the SPR-CNN based channel estimation in the UMi street NLOS scenario, where every $D$ ($=4$) successive coherence intervals are grouped as a CEU. In each CEU, we set $M_{\textrm{T}}[n]=N_{\textrm{T}}=32$, $M_{\textrm{R}}[n]=N_{\textrm{R}}=16$, for $d=1$, and $M_{\textrm{T}}[n]=16$, $M_{\textrm{R}}[n]=4$, for $d=2,3,4$. So the average spatial pilot overhead of the SPR-CNN based approach is $p_{\textrm{SPR-CNN}}=\frac{\sum_{d=1}^{4}p[d]}{4}=88$ and the ratio of it over the full pilot overhead is $r=\frac{p_{\textrm{SPR-CNN}}}{p_{\textrm{full}}}=\frac{88}{256}\approx\frac{1}{3}$. For fair comparison, the ideal and non-ideal MMSE based approaches in Fig.~\ref{nmse_SPR_CNN} also use the above parameter settings. The SF-CNN and SFT-CNN based approaches with full pilot overhead are also plotted for comparison. From the figure, the SPR-CNN based channel estimation achieves comparable performance to the SF-CNN and SFT-CNN based approaches, especially at the low and medium SNR, but only requires about one third of pilot overhead at the cost of complexity. This means that the additional frequency and temporal correlation has been efficiently utilized by the SPR-CNN based approach to reduce the spatial pilot overhead significantly. On the contrary, both the ideal and non-ideal MMSE perform poorly even if using the same channel correlation information as the SPR-CNN based approach, which reveals that the proposed approach is able to tolerate the reduction of spatial pilot overhead.

\section{Conclusion}

In this paper, we have developed the deep CNN based channel estimation approaches for mmWave massive MIMO-OFDM systems. The SF-CNN based channel estimation is first proposed to simultaneously utilize spatial and frequency correlation. To further incorporate the temporal correlation in the real scenario, we develop the SFT-CNN based approach. Finally, considering the huge spatial pilot overhead caused by massive antennas, we design the SPR-CNN based channel estimation to mitigate this problem. Numerical results show that the proposed SF-CNN and SFT-CNN based approaches outperform the non-ideal MMSE estimator yet requiring lower complexity and achieve the performance very close to the ideal MMSE estimator. Even if the channel statistics are different, the proposed approaches can still achieve fairly good performance. The SPR-CNN based channel estimation is efficient to save the spatial pilot overhead significantly with minor performance loss.

The proposed CNN based channel estimation approaches can be further improved in terms of the robustness to as many channel statistics as possible. To achieve this aim, the complicated mmWave MIMO channel structure needs to be deeply dissected to provide insights for the design of DNN architecture and hyper-parameters tuning.

\section*{Acknowledgements}

We would like to thank the support and comments from Intel Corporation and the comments from the reviewers, which have helped us improve the quality of the paper significantly.



\ifCLASSOPTIONcaptionsoff
  \newpage
\fi

\end{document}